\documentclass[aps,prl,twocolumn,reprint,superscriptaddress,showpacs,showkeys]{revtex4-1}
\usepackage{graphicx}
\usepackage{amsmath}
\usepackage{xcolor}
\usepackage{siunitx}

\begin{document}

%\preprint{}

\title{3D strain in 2D materials: Experimental test in unsupported monolayer graphene under pressure}


\author{Y. W. Sun}
\email{yiwei.sun@qmul.ac.uk}
\affiliation{School of Engineering and Materials Science, Queen Mary University of London, London E1 4NS, United Kingdom}
\author{W. Liu}
\thanks{equal contribution to Y. W. Sun}
\affiliation{College of Information Science and Electronic Engineering, Zhejiang University, Hangzhou 310027, China}
\author{I. Hernandez}
\affiliation{Departamento CITIMAC, Universidad de Cantabria, Santander 39005, Spain}
\author{J. Gonzalez}
\affiliation{Departamento CITIMAC, Universidad de Cantabria, Santander 39005, Spain}
\author{F. Rodriguez}
\affiliation{Departamento CITIMAC, Universidad de Cantabria, Santander 39005, Spain}
\author{D. J. Dunstan}
\email{d.dunstan@qmul.ac.uk}
\affiliation{School of Physics and Astronomy, Queen Mary University of London, London E1 4NS, United Kingdom}
\author{C. J. Humphreys}
\email{c.humphreys@qmul.ac.uk}
\affiliation{School of Engineering and Materials Science, Queen Mary University of London, London E1 4NS, United Kingdom}
%\thanks{}


\date{\today}

\begin{abstract}
Previous Raman measurements on supported graphene under high pressure reported a very different shift rate of in-plane phonon frequency of graphene (16 cm$^{-1}$GPa$^{-1}$) from graphite (4.7 cm$^{-1}$GPa$^{-1}$), implying very different in-plane anharmonicity that graphene gets stiffer than graphite in-plane under the same pressure. It was suggested that it could be due to the adhesion of graphene to substrates. We have therefore performed high pressure Raman measurements on unsupported graphene and we find a similar in-plane stiffness and anharmonicity of graphene (5.4 cm$^{-1}$GPa$^{-1}$) to graphite. On the other hand, the out-of-plane stiffness of graphene is hard to define, due to the 2D nature of graphene. However, we estimate a similar out-of-plane stiffness of graphene (1.4$\pm$295 GPa) to that of graphite (38.7$\pm$7 GPa), by measuring its effect on the shift of the in-plane phonon frequency with pressure.  
\end{abstract}

%\pacs{62.50.-p, 63.20.-e, 63.20.dk, 63.22.Np}
%\keywords{}

\maketitle

Monatomic layer materials such as graphene are often described as 2D materials, and indeed their 2D nature has profound dimensionality effects.\cite{Novoselov04,Kawaguchi08,Murphy95} Yet they are often modelled as sheets of isotropic 3D material, with a small effective thickness, even as low as 0.066 nm.\cite{Yakobson96,Wang05} We show here that the true thickness of graphene and its 3D elastic stiffness tensor does retain meaning, corresponding to real experimental observables, in particular that it makes sense to ascribe a 3D strain tensor to monolayer graphene.

Graphene has many extraordinary properties due to its 2D nature. It also brings challenges. For example, a continuum 3D model cannot be applied to graphene to obtain its elasticity. Consequently, the response of graphene (2D) to pressure (3D) cannot be described conventionally. In particular, the out-of-plane stiffness (or the elastic constant c$_{33}$) of graphene is an ambiguous concept and difficult to estimate, due to the lack of definition of out-of-plane strain.

Investigating the mechanical properties of graphene is essential both for fundamental understanding and the development of novel graphene-based nanostructures\cite{Young12} and devices.\cite{Yeh16} The effect of strain on graphene was first studied by Proctor \textit{et al.}.\cite{Proctor09} They performed Raman measurements on graphene on Si/SiO$_2$ substrates under high pressure and reported the shift rate of the in-plane phonon GM frequency with pressure of 16 cm$^{-1}$GPa$^{-1}$, significantly higher than the graphite value of 4.7 cm$^{-1}$GPa$^{-1}$.\cite{Hanfland89} Subsequent work on various substrates (diamond,\cite{Bousige17} sapphire,\cite{Bousige17} copper,\cite{Filintoglou13} and SiO$_2$\cite{Nicolle11}) reports a large range of shift rates from 4.0 to 10.5 cm$^{-1}$GPa$^{-1}$. It is worth noticing that for the measurements on copper, Filingtoglou \textit{et al.} observed a sudden and irreversible change of the GM shift rates from 9.2 to 5.6 cm$^{-1}$GPa$^{-1}$,\cite{Filintoglou13} the latter value being much closer to graphite. They attributed the change to the detachment of graphene from copper. Hadjikhani \textit{et al.} also observed a shift rate of about 5 cm$^{-1}$GPa$^{-1}$ of graphene on copper.\cite{Hadjikhani12} Proctor \textit{et al.} \cite{Proctor09} initially, and Machon \textit{et al.} \cite{Machon17} recently pointed out that the GM shift rate with pressure of supported graphene is determined by substrates via their adhesion to the graphene.

For graphite, we describe the C-C bond stretching by the Morse potential:\cite{Morse29}
\begin{equation}
E(r)=E_0[(1-e^{-\beta(r-r_0)})^2-1],
\label{morse}
\end{equation}
where \textit{r} is the separation of the nearest C-C atoms, $r_0$ is the unstrained C-C bond length, $E_0$ and $\beta$ denote the depth and width of the potential, respectively. The second derivative of \textit{E(r)} gives the force constant \textit{k(r)}, from which we obtain the frequency $\omega(r)$ (cm$^{-1}$), by considering the C-C in-line anti-phase vibration as an harmonic oscillation:
\begin{equation}
\omega(r)=\frac{1}{\pi c}\sqrt{\frac{E_{0}\beta^{2}e^{\beta(r_{0}-r)}(2e^{\beta(r_{0}-r)}-1)}{m}},
\label{omegar}
\end{equation}
where \textit{m} is the mass of a carbon atom and \textit{c} is the speed of light. The C-C separation \textit{r} can be related to the pressure by the 2D in-plane elastic constants $c_{11}^{2D}$ and $c_{12}^{2D}$ (N m$^{-1}$):\cite{Savini11}
\begin{equation}
r(P)=r_{0}(1-\frac{F}{c_{11}^{2D}+c_{12}^{2D}}),
\label{rp}
\end{equation}
where \textit{F} is the in-plane bi-axial force. It is clear that the shift rate of the GM frequency with pressure is determined by the in-plane stiffness:
\begin{equation}
\omega(P)=\frac{1}{\pi c}\sqrt{\frac{E_{0}\beta^{2}e^{\frac{\beta r_{0}a_{33}P}{c_{11}^{2D}+c_{12}^{2D}}}(2e^{\frac{\beta r_{0}a_{33}P}{c_{11}^{2D}+c_{12}^{2D}}}-1)}{m}},
\label{omegap}
\end{equation}
where $a_{33}$ is the interlayer spacing. These equations should apply to unsupported graphene too, except for a caveat on the choice of $a_{33}$. This will be discussed shortly. Proctor \textit{et al.} attempted to make measurements on unsupported graphene, but the specimen also contained multi-layer graphene and nanographite pieces.\cite{Proctor09}

To measure the thickness of one paper, one measures that of a hundred and divides it by 100. Hence it is reasonable to consider the thickness of graphene as 0.34 nm, the interlayer distance in graphite.\cite{Trucano75}. This thickness can be used to obtain the values of the 2D elastic constants and the in-plane bi-axial force in Eq. \ref{rp} for graphene. On the other hand, an effective thickness of graphene is sometimes introduced with an effective Young's modulus, by describing graphene as a continuum plate made of 3D isotropic material,\cite{Yakobson96} for various purposes. For example, Munoz \textit{et al.} described the ballistic thermal conduction of graphene very well by this model.\cite{Munoz10} This approach can lead to a value of thickness as small as 0.066 nm.\cite{Wang05} Despite the success that these definitions bring in many cases, neither definition seems to be appropriate to define the change of thickness resulting from the out-of-plane strain of graphene.

To solve the challenge in estimating the out-of-plane stiffness, we go back to Eq. \ref{omegap} for graphite. We first test how well this theoretical model describes the experimental results by Hanfland \textit{et al.}.\cite{Hanfland89} In our previous work,\cite{Sun13} we obtained the $E_0$ and $\beta$ in Eq. \ref{morse} by \textit{ab initio} calculations,\cite{Holec10} inserted their values and the experimental values for the elastic constants $c_{11}$ and $c_{12}$,\cite{Bosak07} to Eq. \ref{omegap}, and compared the theoretical line for the GM shift rate with pressure $\omega(P)$ to the experimental data. We found that the theoretical line is very straight, not describing the large sublinearity in the experimental data. These two only matched when we introduced an extremely small $c_{33}$ of 39 GPa (compared to 1248 GPa of $c_{11}$+$c_{12}$)\cite{Bosak07} and a relatively large shift rate of $c_{33}$ with pressure, $c_{33}^{\prime}$ at 10,\cite{Gauster74} to the $a_{33}$ in Eq. \ref{omegap}, by $a_{33}=a_{33_{0}}(1+\frac{c_{33}^{\prime}P}{c_{33}})^{-\frac{1}{c_{33}^{\prime}}}$, where $a_{33_{0}}$ is the unstrained interlayer distance. This indicates that the very soft out-of-plane stiffness of graphite, along with its relatively large shift rate with pressure, are responsible for the sublinearity in the shift of the in-plane phonon (GM) frequency of graphite with pressure. We now use this approach to estimate the out-of-plane stiffness of graphene.

In this work, we apply high pressure to unsupported graphene in N,N-Dimethylformamide (DMF), measure the in-plane stiffness of graphene, and estimate the out-of-plane stiffness by its effect on the shift of the in-plane phonon frequency with pressure.

We used chemical vapor deposition (CVD) monolayer graphene grown on copper. We took advantage of a wet transfer method\cite{Li09} to obtain unsupported monolayer graphene in solution, as briefly described below. First, a thin poly(methyl methacrylate) (PMMA) layer was spin coated on graphene/Cu and baked. Then we removed the substrate copper in an etchant (CuSO$_4$$\cdot$5H$_2$O and HCl), leaving PMMA/graphene floating on the surface of the solvent. We transferred the PMMA/graphene membrane into DMF after rinsing the membrane in deionized water. The graphene was free-floating in DMF as the covering PMMA had been dissolved. As we know there is no stable graphene suspension, so the free-floating graphene should slowly precipitate and it is difficult to locate. In order to get a good signal from the Raman measurements, we increased the concentration of graphene by keeping transferring PMMA/graphene membranes into DMF until saturated PMMA appeared. We gently heated the solution at \SI{50}{\celsius} to further increase the concentration of graphene and took some of the solution from the bottom, as sample D, graphene in DMF.  We took the saturated PMMA (still covering the graphene) out of the solution as sample P, graphene in PMMA. We performed Raman measurements under high pressure on both samples.

To apply high pressure, we loaded the samples into a membrane diamond anvil cell with anvils of 500 $\mu$m culet size and Type \uppercase\expandafter{\romannumeral 2}a diamonds. The ruby luminescence R1 line was used for pressure calibration.\cite{Mao86} DMF and PMMA are not common pressure transmitting media (PTM) in high pressure experiments as the hydrostaticity is not guaranteed. We loaded 3 ruby pieces into the cell in different locations to monitor the hydrostaticity at high pressure. The full width at half maximum (FWHM) of the GM Raman peaks of graphene, plotted with pressure in the supporting information (SI), can be used as an additional indication of hydrostaticity. We stopped increasing the pressure when a large non-hydrostaticity was observed at 5 GPa for sample P and 7 GPa for sample D.

We performed non-polarized Raman measurements in the backscattering geometry with a Horiba T64000 Raman system with a confocal microscope, a single 1800 grooves/mm grating, a 100 $\mu$m slit and a liquid N$_2$-cooled CCD detector (Jobin-Yvon Symphony). For both samples, the highest signal to noise ratio of the Raman signal is obtained at 514 nm excitation, compared to 488 and 647 nm, at the same spectra-collecting condition. An edge filter for the 514 nm line from a Coherent Innova Spectrum 70C Ar$^{+}$-Kr$^{+}$ laser was used. We kept the laser power on the sample below 5 mW to avoid significant laser-heating effects on the graphene and the concomitant softening of the Raman peaks.

\begin{figure}
\includegraphics[width=0.9\columnwidth]{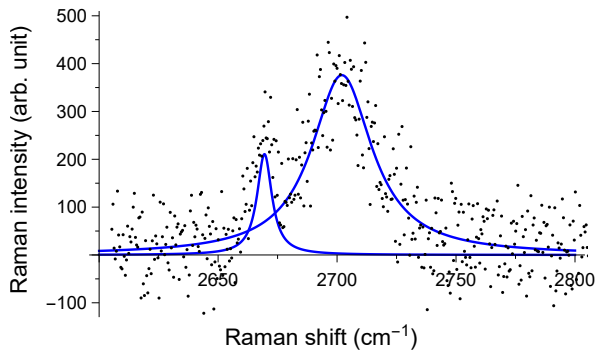}
\caption{Background-subtracted spectrum (black dots) of sample D at the first pressure point is plotted over the 2D range. The optimum fit to two Lorentzians is shown as blue lines.}
\label{eg}
\end{figure}

We sought to obtain unsupported monolayer graphene in DMF. The free-floating monolayers may interact by van der Waals bonding to form multi-layer graphene --- more likely when the graphene concentration increases. We observed a single and strong 2D peak for sample P, having a 2D-to-G integrated area ratio of 1.3, similar to that of supported monolayer graphene. For sample D, however, the 2D peak is weaker than the G. It is worth noticing that many factors determine the intensity of a Raman peak and there is no previous observation on the 2D peak of unsupported graphene. The number of 2D peaks, on the other hand, depends on the interlayer interaction between graphene layers, via the double resonance process.\cite{Ferrari06} To obtain the number of peaks in a spectral profile, we employed maximum likelihood estimation to fit the spectrum and compare different fits by an objective criterion, the Bayesian Information Criterion (BIC).\cite{Schwarz78} Briefly, an increasing number of peaks (therefore fitting parameters) usually brings the fit closer to the data. The BIC gives a penalty to introducing additional fitting parameters to avoid over-fitting. We present the best fit of the spectrum of sample D at the first pressure point over the 2D range after subtracting the background by a 5-term polynomial in Fig. \ref{eg}. The fit consists of two Lorentzians --- the left from the diamond and the other from graphene. Increasing the number of peaks or changing the shape of the peaks increases the BIC, and is therefore a less good fit to the data. We reasonably conclude that the interaction between the graphene pieces in sample D is too weak to form multilayer graphene from the objectively fitted single 2D peak.

\begin{figure}
	\includegraphics[width=\columnwidth]{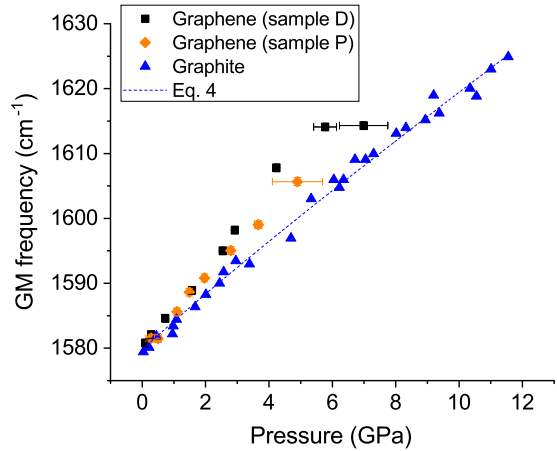}
	\caption{The GM frequency of graphene in samples D (black squares) and P (orange diamonds) is plotted versus pressure. The data of graphite\cite{Hanfland89} (blue triangles) are also plotted with the theoretical line of Eq. \ref{omegap} (blue dashed line) for comparison. The uncertainty in both pressure and frequency is plotted where it exceeds the size of a data point.}
	\label{data}
\end{figure}

We move on to the response of unsupported graphene to pressure. We present all the GM spectra under pressure in the SI. Fig. \ref{data} shows the frequency of the in-plane phonon GM of graphene in samples P and D at various pressures, to compare with graphite data.\cite{Hanfland89} The theoretical line of Eq. \ref{omegap} is also plotted and describes the graphite data very well. The uncertainty in frequency comes from the fitting and the resolution of the Raman system. It is very small and barely exceeds the size of a data point for all the graphene data. The uncertainty in pressure comes from the measurements on different ruby pieces, the larger R-line deviation, the higher non-hydrostaticity. We rule out the last two data points of sample D and the first and the last points of sample P in the following fitting as they are clearly not under good hydrostatic conditions. In general, unsupported graphene behaves very similarly to graphite under pressure, in terms of the shift rates of the GM, in contrast to previous published results on supported graphene.

Sample D contains unsupported monolayer graphene in liquid solution. We have 7 data points under reasonably good hydrostatic condition. It appears that the first 4 and the last 3 behave differently. To objectively determine it, we linear-fit all the 7 as model 1 and separately linear-fit the first 4 and the last 3 as model 2. We again employ the maximum likelihood estimation and compare the corrected Akaike information criterion (AICc) (used when the number of data points is small) of these two models,\cite{Cavanaugh97} and find that the separate fit is significantly preferred by the data (21.55 vs. 28.42 of AICc). It is reasonable to attribute the first 4 data points to unsupported graphene, with a slope of 5.4 cm$^{-1}$GPa$^{-1}$, very close to 4.7 cm$^{-1}$GPa$^{-1}$ of graphite. We think that the higher slope of the last 3 at 7.5 cm$^{-1}$GPa$^{-1}$ is likely due to the adhesion to the surrounding solidified DMF, similar to supported graphene on a substrate. 

\begin{figure}
	\includegraphics[width=0.9\columnwidth]{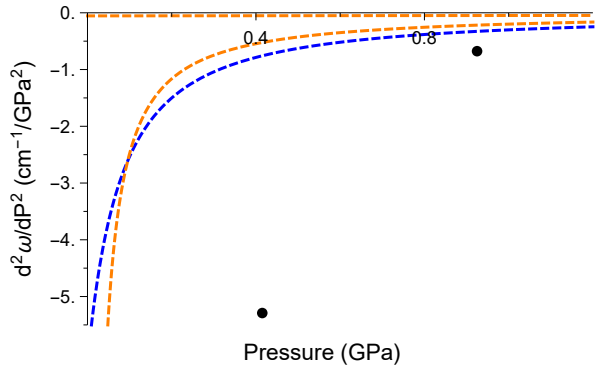}
	\caption{The second derivative of the GM frequency to pressure (black dots) of the graphene in sample D is plotted versus pressure, with the optimal fit (blue dashed line) by the second derivative of the Eq. \ref{omegap} to pressure. Two curves (orange dashed lines) by increasing or decreasing the value of $c_{33}$ by 100 times are plotted to be compared with the optimal fit.}
	\label{dmfs}
\end{figure}

\begin{figure}
	\includegraphics[width=0.9\columnwidth]{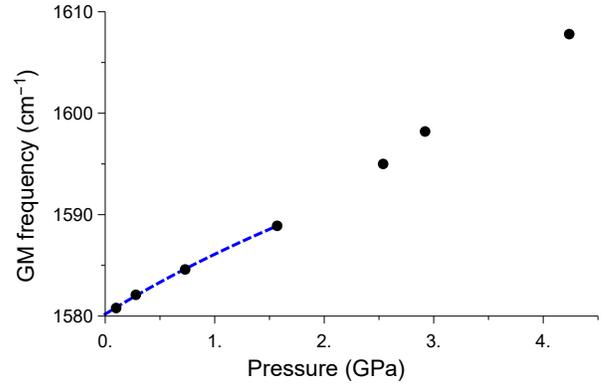}
	\caption{The GM frequency of the graphene in sample D is plotted versus pressure. These data points, taken from Fig. \ref{data}, have a small uncertainty in pressure, implying good hydrostaticity. The optimal empirical fit of the first four points (unsupported graphene) is shown in blue dashed line.}
	\label{dmf}
\end{figure}

We now focus on the sublinearity of the GM shift with pressure. As we mentioned, the curvature is due to the large reduction of the in-plane bi-axial force from the large anisotropy of graphite. We attempt to apply Eq. \ref{omegap} to the first 4 points. Despite the shift rates being similar for graphite (4.7 cm$^{-1}$GPa$^{-1}$) and graphene (5.4 cm$^{-1}$GPa$^{-1}$), this difference is much larger than the difference in the frequency at zero pressure, 1578.8 cm$^{-1}$ for graphite and 1580.2 cm$^{-1}$ for graphene. While in Eq. \ref{omegap} the intercept ($\propto\sqrt{E_0}\beta$) and slope ($\propto\sqrt{E_{0}\beta}$) are determined by the same factors, the best fit of the graphene data (sample D and P) will give a lower slope than required to keep the intercept and therefore the whole fitting optimal. Consequently, $c_{33}$ and $c_{33}^\prime$ can only make the fitting curve straight (by being infinitely large) to compensate the lower initial slope, and will not be able to describe the curvature. On the other hand, we have 4 data points, just more than enough to determine the second derivative of Eq. \ref{omegap} to pressure, which extracts the curvature regardless of the slope. We keep the value of all the parameters the same as graphite, except $c_{33}$ as the fitting parameter and obtain its optimal value at 1.4$\pm$295 GPa, compared to 38.7$\pm$7 GPa of graphite.\cite{Bosak07} Despite the large error, $c_{33}$ is much smaller than $c_{11}+c_{12}$. We present the optimal fit of the second derivative of the Eq. \ref{omegap} with the data in Fig. \ref{dmfs}. Alternatively, since we know that the curvature is determined by the out-of-plane stiffness ($a_{33}=a_{33_{0}}(1+\frac{c_{33}^{\prime}P}{c_{33}})^{-\frac{1}{c_{33}^{\prime}}}$), we can empirically fit the data of graphite and graphene by $\omega=\omega_0(\delta P+1)^{\delta^{\prime}}$ with $\delta^{\prime}\propto 1/c_{33}^{\prime}$ and $\delta^{\prime}\delta\propto 1/c_{33}$. From the optimal fit, $\delta^{\prime}$ is 0.054$\pm$0.01 for graphite and 0.014$\pm$0.006 for graphene. $\delta^{\prime}\delta$ is (3.2$\pm$1.4)$\times$10$^{-3}$ GPa$^{-1}$ for graphite and (4.3$\pm$3.9)$\times$10$^{-3}$ GPa$^{-1}$. We present all the data points at the hydrostatic condition of sample D and the optimal empirical fit of the first 4 points in Fig. \ref{dmf}. The results suggest that unsupported graphene in solution presents a similar out-of-plane stiffness to graphite, as expected.

\begin{figure}
	\includegraphics[width=0.9\columnwidth]{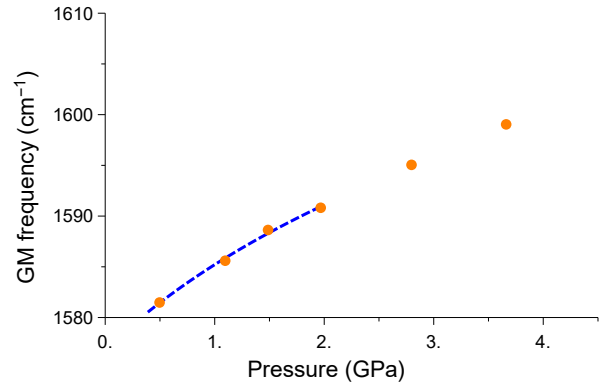}
	\caption{The GM frequency of the graphene in sample P is plotted versus pressure. These data points, taken from Fig. \ref{data}, have small uncertainties in pressure, implying good hydrostaticity. The fit of the first four points over the similar range to the fit in Fig.\ref{dmf} is shown in blue dashed line.}
	\label{pmma}
\end{figure}

The PMMA used in this study is a gel. The adhesion of the gel to graphene is unclear. We use sample P in comparison to sample D and therefore we apply the same empirical fitting as before to the data over a similar pressure range, after ruling out those at non-hydrostatic conditions. The slope is similar to sample D (see Fig. \ref{data}) and the fitting results for the curvature are similar, giving $\delta^{\prime}$ of 0.013$\pm$0.01 and $\delta^{\prime}\delta$ of 6.6$\pm$14.8 GPa$^{-1}$. This indicates that the gel, like a liquid, and unlike solid substrates, has little adhesion to graphene under pressure. Then we can consider the graphene in sample P as `unsupported' and it again shows a similar both in-plane and out-of-plane stiffness to graphite. We present the data and corresponding fit in Fig. \ref{pmma}.

In conclusion, we performed Raman measurements on unsupported graphene in DMF and PMMA under pressure. We find that the shift rate of the in-plane phonon GM frequency of graphene (5.4 cm$^{-1}$GPa$^{-1}$) with pressure is close to that of graphite (4.7 cm$^{-1}$GPa$^{-1}$), in contrast to previous high-pressure measurements on supported graphene (16 cm$^{-1}$GPa$^{-1}$). Our results indicate a similar in-plane stiffness and anharmonicity of graphene to graphite, again in contrast to previous work. The small out-of-plane stiffness of graphite results in a reduction of the in-plane force under pressure, and therefore the GM frequency shifts sublinearly with pressure. We estimate a similar out-of-plane stiffness for graphene	(1.4$\pm$295 GPa) to graphite (38.7$\pm$7 GPa) from this effect and we consider that this is a reliable and meaningful way to estimate the out-of-plane strain and stiffness for 2D materials, as it corresponds to real experimental observables and does not involve ambiguously defined, physically meaningless effective thickness.

\begin{acknowledgments}
The authors are grateful for valuable comments on the interpretation of the results by Dr. D. Holec from Montanuniversitat Leoben and Prof. A. San Miguel from Univ Lyon 1. 
\end{acknowledgments}
\bibliography{apssamp1}

\end{document}